\documentclass{aa}
\usepackage{graphics}
\usepackage{amssymb}

\begin{document}

\title{Investigation of the Neupert effect in solar flares.
I. Statistical properties and the evaporation model}
\titlerunning{The Neupert effect -- statistical properties}

\author{A.~Veronig\inst{1}
 \and B.~Vr\v{s}nak\inst{2}
 \and B. R.~Dennis\inst{3}
 \and M.~Temmer\inst{1}
 \and A.~Hanslmeier\inst{1}
 \and J.~Magdaleni\'{c}\inst{2}
 }

\offprints{A. Veronig} \mail{asv@igam.uni-graz.at}

\institute{Institute for Geophysics, Astrophysics and Meteorology,
University of Graz, Universit\"atsplatz 5, A-8010 Graz, Austria
\and Hvar Observatory, Faculty of Geodesy, University of Zagreb,
 Ka\v{c}i\'ceva 26, HR-10000 Zagreb, Croatia
\and NASA Goddard Space Flight Center, Greenbelt, MD 20771, U.S.A.
 }

\date{Received xxxx/ Accepted xxxx}

\abstract{Based on a sample of  1114 flares observed simultaneously in
hard X-rays (HXR) by the BATSE instrument and in soft X-rays (SXR) by
GOES, we studied several aspects of the Neupert effect and its interpretation in the
frame of the electron-beam-driven evaporation model. In particular,
we investigated the time differences ($\Delta t$) between the maximum of the
SXR emission and the end of the HXR emission, which are expected to occur at
almost the same time. Furthermore, we performed a detailed analysis of the SXR peak
flux -- HXR fluence relationship for the complete set of events, as well as separately
for subsets of events which are likely compatible/incompatible with the timing
expectations of the Neupert effect. The distribution of the time
differences reveals a pronounced peak at $\Delta t = 0$. About half of the events
show a timing behavior which can be considered to be consistent with the
expectations from the Neupert effect. For these events, a high correlation between
the SXR peak flux and the HXR fluence is obtained, indicative of
electron-beam-driven evaporation. However, there is also a significant
fraction of flares (about one fourth), which show strong deviations from
$\Delta t = 0$, with a prolonged increase of the SXR emission distinctly
beyond the end of the  HXR emission. These results suggest that electron-beam-driven
evaporation plays an important role in solar flares. Yet, in a significant fraction
of events, there is also clear evidence for the presence of an additional
energy transport mechanism other than nonthermal electron beams,
where the relative contribution is found to vary with the flare importance.
 }

\maketitle

\section{Introduction}

Observations of solar flares in X-rays and microwaves frequently
show that the shape of the rising part of the soft X-ray light
curve closely resembles the time integral of the microwave or hard
X-ray light curve. This led to the idea that there is a causal
relationship between the nonthermal (microwave and hard X-ray) and
thermal (soft X-ray) emission of a flare (Neupert 1968; Dennis \&
Zarro 1993), which has become known as the Neupert effect.

It has been shown that this effect can be reproduced by a model,
in which the flare energy is released primarily in the form of
nonthermal electrons (e.g., Brown 1971; Li et al. 1993). According
to the so-called thick-target model, the hard X-ray (HXR) emission
is electron-ion bremsstrahlung produced by electron beams
encountering the dense layers of the lower corona, transition region, and
chromosphere. The model assumes that only a small fraction of the
energy of the nonthermal electrons is lost through radiation (for
a discussion see McDonald et al. 1999). Most of the energy is
transferred to heating of the ambient thick-target plasma via
Coulomb collisions between the beam and the ambient electrons. Due
to the rapid deposition of energy by the accelerated electrons,
the energy cannot be radiated away at a sufficiently high rate and
strong pressure gradients develop. The pre-flare hydrostatic
equilibrium is lost and the heated plasma explosively expands up
into the corona in a process known as chromospheric evaporation
(e.g., Antonucci et al. 1984; Fisher et al. 1985; see also the
review by Antonucci et al. 1999, and references therein). The hot
dense plasma that has been convected into the corona gives rise to
enhanced soft X-ray (SXR) emission via thermal bremsstrahlung.

Under such circumstances, the hard X-ray emission is directly
related to the electron beam flux. On the other hand, the soft
X-ray emission should be directly linked to
the energy deposited by the same electrons up to a given time,
i.e. to the time integral of the electron beam flux,
and we can expect to see the Neupert effect. Thus, if
the Neupert effect is observed, this can be considered as
\mbox{evidence} of electron-beam-driven chromospheric evaporation (see
McTiernan et al. 1999). In recent years, the Neupert effect has
also been observed on stars, which suggests the existence of
the chromospheric evaporation process also in stellar flares
(Hawley et al. 1995; G\"udel et al. 1996).

In the present study we utilize statistical properties of solar
flares observed simultaneously in SXR and HXR emission to test several
expectations from the Neupert effect. The main predictions are:
(1) The maximum of the SXR emission and the end of the HXR
emission should occur at the same time. (2) There should be a
high correlation between the HXR fluence, i.e. the HXR flux
integrated over the event duration, and the SXR peak flux. The
correlation of the HXR fluence and the SXR peak flux and its
relation to the involved nonthermal and thermal energies provide
the fundamental link between the Neupert effect and the
electron-beam-driven chromospheric evaporation model, which will be discussed in
Sect.~\ref{SectAnalysis}. The present analysis can be considered
as a complementary approach to studies of the Neupert effect which
use the actual SXR and HXR light curves (see, e.g., Dennis et
al. 1992; Dennis \& Zarro 1993; McTiernan et al. 1999).
By accessing only statistical flare
quantities, such as HXR end time, SXR maximum time, HXR fluence
and SXR peak flux, we neglect part of the information contained in
the light curves. However, such a statistical approach has the
advantage that it can be applied to a large data set, i.e. it
is not restricted to a selected sample of events, which intrinsically
favors the analysis of large flares. Moreover, it enables us to
define and investigate subsets of events, still representing
statistically meaningful data sets.

The paper is structured in the following way. Section~2 contains a
description of the soft X-ray and hard X-ray data used in the analysis
together with the method of finding corresponding SXR/HXR events.
In Section~3, we discuss, in the frame of the electron-beam-driven chromospheric
evaporation model, the relationship between SXR and HXR emissions, and the
associated thermal/nonthermal energies. In this respect, it is
essential to clarify the question {\it if\,} and {\it in which
formulation} the Neupert effect is valid for the bulk of solar flares.
In Section~4, our results are presented, comprising an investigation
of the relative timing of the SXR and HXR emission as well as a
detailed study of the HXR fluence~-- SXR peak flux relationship. The
results are interpreted and discussed in Section~5, and the conclusions
are drawn in Section~6.

\section{Data set}

In the present study the soft X-ray and hard X-ray bursts are
compared using the SXR data from the {\it Geostationary
Operational Environmental Satellites} (GOES) and the HXR data
from the {\it Burst and Transient Source Experiment}
(BATSE) aboard the {\it Compton Gamma-Ray Observatory} (CGRO).
The X-ray sensor aboard GOES consists of two ion chamber detectors,
which provide whole-sun X-ray fluxes in the 0.05--0.4 and
0.1--0.8~nm wavelength bands. A description of the GOES ion
chambers can be found in Donelly \& Unzicker (1974) and Garcia
(1994). BATSE is a whole-sky HXR flux monitor that consists of
eight large-area wide-field detectors, placed on the corners of
the CGRO spacecraft. From each of eight detectors there are hard
X-ray data in four energy channels, 25--50, 50--100, 100--300 and
$>$300~keV, obtained with a time resolution of about 1~s.
Technical characteristics of the BATSE instrument and its
application to solar flare studies are described in Fishman et al.
(1989, 1992) and Schwartz et al. (1992).

We utilize the 1-min averaged GOES soft X-ray data in
the 0.1--0.8~nm wavelength band as listed in the flare compilation of the Solar Geophysical
Data (SGD, {\tt \small ftp://ftp.ngdc.noaa.gov/STP/SOLAR$\_$DATA/SOLAR$\_$FLARES/}),
and the hard X-ray data from the BATSE Solar Flare Catalog, archived in the
Solar Data Analysis Center (SDAC) at NASA/Goddard Space Flight Center
(GSFC, {\tt \small ftp://umbra.nascom.nasa.gov/pub/batse/}).
In the BATSE Flare Catalog the start, maximum, and end time
of an event are listed with an accuracy of
1~s. The peak count rate and the total count rates are
background subtracted for the flux below 100~keV. For
the SXR events observed by GOES we used the flux just before the
flare start for the background subtraction. The analysis
was carried out for the period from January 1997 to June 2000 (when CGRO
was deorbited), for which 6947 SXR events and 2738 HXR events
are reported.

Due to the lack of spatial information, the
determination of related SXR/HXR flares is exclusively based
on temporal coincidence. To be identified as corresponding events,
we demand that the start time difference between a SXR and a HXR
event does not exceed 10~min. To avoid as much as possible
any incidental assignment, we applied the following refinements.
All SXR (HXR) events that overlap in time
with any other SXR (HXR) event are excluded. Moreover, events
for which a multiple assignment is possible (e.g., one SXR event
can be related to two different HXR events by the start time
criterion) are excluded from the analysis. Applying these
criteria, we obtained 1404 events that were observed in both SXR
and HXR emissions. This data set was reduced to a final set of
1114 SXR/HXR events, after excluding those events with missing
SXR background flux data.

The characteristic times, which are under study,
are the peak time of the SXR emission and the end time of the HXR
emission. It has to be noted that the end of a HXR event is difficult
to determine, whereas the SXR peak time is a rather well defined quantity
(at least within the given precision of 1~min). So, it has to be
kept in mind that there is a statistical error, in particular related
to the HXR end time, which introduces also a statistical error on the
HXR fluence data. Thus, a scatter in the relevant figures has
to be expected in addition to the scatter caused by different
physical conditions in flares. Moreover, we stress that for flares
of low SXR or low HXR intensities, the peak flux and fluence data
are affected by observational selection/sensitivity effects, which has to
be considered when interpreting the data. In each of the figures, the
estimated threshold ranges for the relevant parameters are indicated.
See Lee et al. (1993) for methods of handling these selection
effects.

\section{The Neupert effect and the chromospheric evaporation model
\label{SectAnalysis} }

The Neupert effect, as it is commonly stated in the literature,
can be expressed as (e.g., Lee et al. 1995)
\begin{equation}
F_{\rm P,SXR} = k \cdot {\cal F}_{\rm HXR}  , \label{EqNeup}
\end{equation}
whereas $F_{\rm P,SXR}$ denotes the SXR peak flux, and
\begin{equation}
{\cal F}_{\rm HXR} = \int_{t_0}^{t_0+D} F_{\rm HXR}(t)dt
\label{EqFluence}
\end{equation}
the HXR fluence, i.e. the HXR flux, $F_{\rm HXR}(t)$, integrated
over the event duration~$D$ starting at time~$t_0$.
The coefficient~$k$ depends on several factors,
as, for instance, the magnetic field geometry and the viewing angle, and
thus may vary from flare to flare (see Lee et al. 1995). However,
if $k$ does not depend systematically on the flare intensity, then
from the Neupert effect stated in Eq.~(\ref{EqNeup}), a linear
relationship is expected to exist between the SXR peak flux
and the HXR fluence.

On the basis of flare frequency distributions, Lee et al. (1993, 1995)
and Veronig et al. (2002a) found inconsistencies with the
linearly formulated Neupert effect as given in Eq.~(\ref{EqNeup}).
If $k$ does not depend systematically
on the flare intensity, then the HXR fluence and the SXR peak flux
distributions should have the same shape, in particular they
should be described by the same power-law index. However, the
power-law index derived from HXR fluence distributions, $1.4
\lesssim \alpha \lesssim 1.6$ (cf. Lee et al. 1993, and references
therein) is distinctly smaller than those obtained for SXR peak
flux distributions, $1.8 \lesssim \alpha \lesssim 2.0$ (cf.
Veronig et al. 2002a, and references therein).

A possible explanation for this discrepancy is that the HXR and
SXR emissions are not directly indicative for the energies
involved, in the sense that the energies are not simply linearly
related to the emissions. As emphasized in Lee et al. (1995), the
Neupert effect interpreted as a consequence of electron-beam-driven
chromospheric evaporation, should exist not necessarily between
the X-ray emissions but between the energies. In the frame of the
electron-beam-driven evaporation model, the HXR emission is a measure of
the rate of energy deposition by accelerated electrons and the SXR
emission is a measure of the total energy contained in the plasma
heated by thermalization of the same electrons. Thus, the energy
deposited by the nonthermal electrons, $\epsilon_{{\rm e}^-}$,
should be equal to the maximum thermal energy contained in the
plasma that is heated by this electron population, $\epsilon_{\rm
th,max}$, i.e.
\begin{equation}
\epsilon_{{\rm e}^-} = \epsilon_{\rm th,max} \, . \label{NeupEn}
\end{equation}

Since the relation between the energies and the X-ray emissions
is not necessarily linear, Eq.~(\ref{NeupEn}) is compatible
with the possibility that the factor~$k$ may be a function of the flare
intensity, violating a linear interpretation of the Neupert effect formulated
for the X-ray emissions (Eq.~(\ref{EqNeup})). Thus, a dependence of $k$ on
the flare intensity does not necessarily indicate that the Neupert
effect formulated for the energies (Eq.~(\ref{NeupEn})), i.e. the
electron-beam-driven chromospheric evaporation model, is violated. Results
reported in several recent papers indicate that~$k$ indeed might
depend on the flare intensity, suggesting that the amount of SXR
emission per HXR electron may differ for small and large flares
(Jim McTiernan, private communication). Feldman et al. (1996) and
Garcia (2000) report that the observed SXR temperature tends to
increase with flare intensity. On the other hand, as shown by McTiernan
et al. (1999), consistency of the observed HXR and SXR emission with
the Neupert effect depends on the temperature response of the SXR
detector used. The Neupert effect is more commonly associated with
SXR emission at high than low temperatures (McTiernan et al.
1999), which might be a further indication for an interdependence
of $k$ with the flare intensity. Furthermore, Tomczak (1999) reports
that the photon spectral index of the measured HXR emission as well
as the relative SXR -- HXR productivity depend systematically on the
flare intensity. However, no systematic dependence of the photon
spectral index on the flare intensity was found by Dennis (1985),
especially for gradual flares.

Another possibility is that the Neupert effect,
formulated for the X-ray emissions (Eq.~(\ref{EqNeup})) as well as
for the more fundamental relationship between the thermal and
nonthermal energies (Eq.~(\ref{NeupEn})), does not hold for the
bulk of flares but maybe only for a specific subset. Most
observational evidence for the Neupert effect is provided for
large and impulsive flares (e.g., Dennis \& Zarro 1993; McTiernan
1999). Any deviation from the Neupert effect, in principle, means
that the hot SXR emitting plasma is not heated exclusively by
thermalization of the accelerated electrons that are responsible
for the HXR emission (Dennis \& Zarro 1993; Lee et al. 1993). In
this case, an additional energy term has to be added on the left
hand side of Eq.~(\ref{NeupEn}).

Several attempts have been made to investigate the relationship
between energies associated with HXR and SXR bursts estimating
the total energy contained in precipitating electrons from
measured HXR spectra and the thermal energy of the heated plasma
from SXR measurements (e.g., Tanaka et al. 1982; Antonucci et al.
1984; Wu et al. 1986; Starr et al. 1988). However, as pointed out
by Wu et al. (1986), neither the thermal nor the nonthermal energy
can be estimated to better than an order of magnitude. The
uncertainties of the nonthermal energy calculations from the HXR
measurements are basically due to the fact that the low energy
cut-off in the electron spectrum is unknown and may vary from
flare to flare. Furthermore, the low energy cut-off may also vary
during a given flare (Gan et al. 2002). The thermal energy
calculations are uncertain primarily due to the estimates of the
volume, filling factor and density of the emitting plasma.

In the following, we analyze the Neupert effect comparing the
directly observable X-ray emissions. One aspect is the analysis of
the relative timing of corresponding SXR and HXR events. The
interpretation of the Neupert effect in the frame of the
electron-beam-driven evaporation model implies that the end of the HXR
burst should be coincident with the maximum of the SXR
emission: When the electron input stops, the HXR emission also has
to stop, and the SXR emission does not further increase. In
principle, the HXR and SXR emission should also start at the same
time. However, in more than 90\% of flares the SXR emission starts
before the HXR emission by at least several minutes (Veronig et al. 2002b).
This may be indicative of a thermal preheating phase prior to the
impulsive electron acceleration or it may be related to the sensitivity
threshold of the hard X-ray detectors (see also Dennis 1988). Thus,
we do not incorporate an investigation of the HXR~-- SXR start time
differences.

The other aspect is to analyze the HXR fluence~-- SXR peak flux
relationship. A simple prediction of the Neupert effect is that
there should be a high correlation among these two parameters,
even if the relationship is not linear. In particular, we
will also investigate the factor~$k$ as a function of the flare
intensity. Additionally, making use of the results from the timing
analysis, different subsets of events will be extracted, which are
likely compatible/incompatible with the Neupert effect regarding
their temporal behavior. Differences in the HXR fluence~-- SXR
peak flux relationship between these subsets may help in understanding
the role of~$k$ in the frame of the Neupert effect.

\section{Analysis and Results}

\subsection{SXR -- HXR timing}

For each event we determined the difference of the peak time of the SXR
emission,
$t_{\rm SXR,P}$, and the end time of the HXR emission, $t_{\rm HXR,E}$:
\begin{equation}
\Delta t = t_{\rm SXR,P} - t_{\rm HXR,E} \, .
\end{equation}
Furthermore, the time differences were normalized to the
duration~$D$ of the respective HXR event:
\begin{equation}
\Delta t_{\rm norm} = \frac{\Delta t}{D} \, .
\end{equation}
The normalized time differences are of particular interest when
the timing behavior of long-duration flares is considered. Such
events may show considerable time differences but these may be
small compared to the overall duration of the event. As intense flares tend to be of
longer duration than weak flares (see e.g., Crosby et al. 1998; Veronig et
al. 2002a), applying a criterion exclusively based on absolute time
differences will act selectively on intense flares. From the present
data set we obtain a cross-correlation coefficient (calculated in
logarithmic space), $r = 0.47$ for the SXR peak flux and SXR flare
duration, and $r = 0.55$ for the HXR peak flux and HXR event duration,
indicating a distinct correlation between the intensity of an event
and its endurance. In general, the duration of the HXR emission of a
flare is much shorter than that of the SXR emission. From the present
data set, we derive a median duration (given with 95\% confidence interval)
of $1.9\pm 0.2$~min for the HXR events and $12.0\pm 0.5$~min for the SXR events.

\begin{figure}
\centering
\hspace*{-0.2cm}
\resizebox{0.95\hsize}{!}{\includegraphics{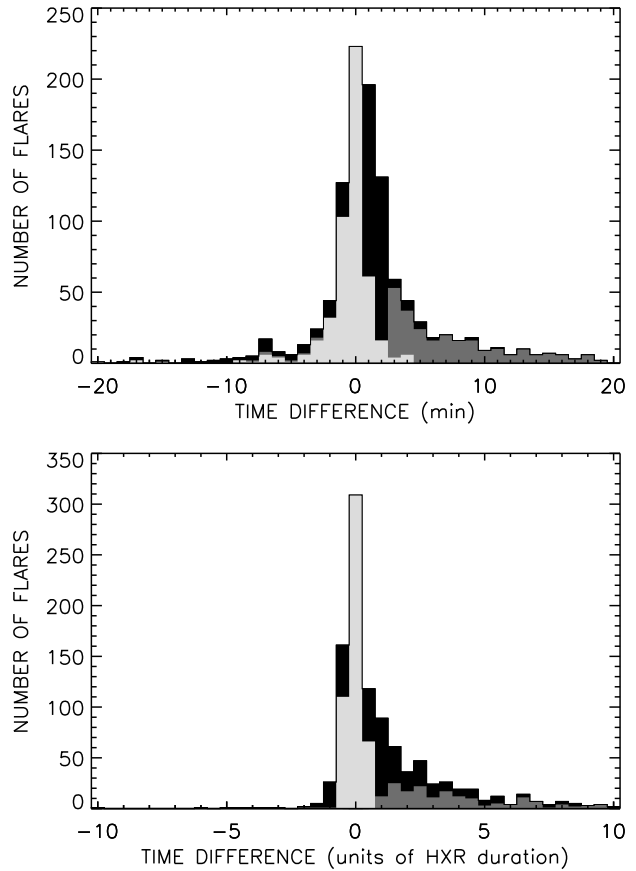}}
    \caption{Histogram of the difference of the SXR maximum and HXR end time,
    given in absolute values (top panel) and normalized to the
    HXR event duration (bottom panel). Positive values indicate that the
    maximum of the SXR emission occurs after the end of the HXR emission,
    negative values vice versa. The shading refers to different
    samples of events, which are compatible with the timing
    expectations of the Neupert effect (light grey, set~1),
    strongly incompatible (dark grey, set~2), or lie in between these
    two extremes (black). For further explanations see
    Sect.~\ref{Sect_combi}. \label{Fig_timing} }
\end{figure}

Figure~\ref{Fig_timing} shows the histogram of time differences derived
for 1114 SXR/HXR bursts, in absolute values (top panel) and normalized to
the HXR event duration (bottom panel). As the HXR times are given with an accuracy of
1~s and the SXR times with an accuracy of 1~min, we cannot expect to obtain
reliable time differences $\lesssim$\,1~min. Thus, for
the derivation of the time differences, the HXR times have been rounded to
minutes. The histogram of the $\Delta t$ uses a bin size of 1~min.
In the histogram of the $\Delta t_{\rm norm}$, a bin size
of 0.5 units of the HXR duration is used. The HXR events have a median duration
of 1.9~min, thus on average a time difference of 0.5~units in the normalized
representation can be considered to roughly correspond to a time difference
of 1~min for the absolute values.

Figure~\ref{Fig_timing} clearly reveals that both representations
of the SXR -- HXR time difference have its mode at zero.
Almost half of the events (49\%) lie within the range $|\Delta t| \leq 1$~min,
and 65\% within the range $|\Delta t| \leq 2$~min.
For the normalized differences we obtain that 44\% lie within the
range $|\Delta t_{\rm norm}| \leq 0.5$~units, and 59\% within
$|\Delta t_{\rm norm}| \leq 1$~unit.
This outcome suggests that certainly a considerable part
of the events shows a good agreement with the expectations from
the Neupert effect regarding the relative timing of the SXR and
HXR emission.
Furthermore, the histograms in Fig.~\ref{Fig_timing} show that there
are more events, for which the SXR maximum takes place after the HXR end
(56\%) than vice versa (24\%). 20\% of the events do not
show a distinguishable time difference, i.e. the SXR maximum and
the HXR end take place within 1~min. This asymmetric behavior is
particularly evident for the distribution of the normalized time
differences.

\begin{figure}
\centering
\hspace*{-0.2cm}
\resizebox{0.97\hsize}{!}{\includegraphics{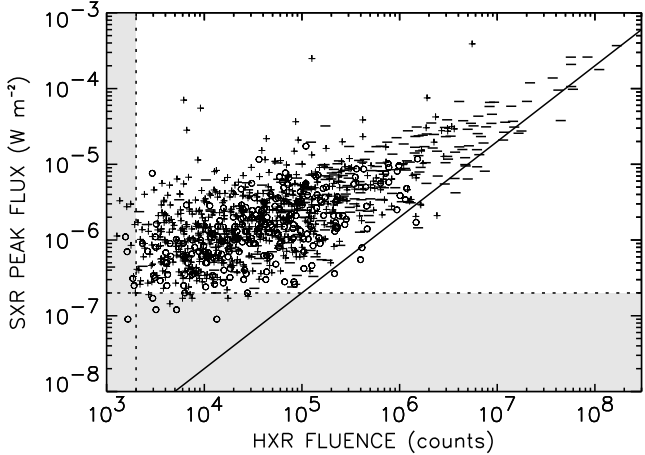}}
    \caption{Scatter plot of the SXR peak flux versus the HXR fluence
    for the complete sample. The vertical dashed line
    indicates the HXR fluence threshold, the horizontal dashed
    line the SXR peak flux threshold. The regions that lie outside the
    respective detection limits are grey shaded. Moreover, we have indicated
    the sign of the time difference between SXR peak and HXR end
    for each single event: ``$+$" symbols denote events with positive,
    ``$-$" symbols events with negative, ``$\circ$" symbols events with zero
    time difference. The straight line indicates a line of
    constant $k$, i.e.\ $F_{\rm P,SXR} = 2\cdot 10^{-12} \cdot {\cal F}_{\rm HXR}$.
    \label{Fig_FluencePeak} }
\end{figure}

\subsection{SXR peak flux -- HXR fluence relationship}

Figure~\ref{Fig_FluencePeak} shows the scatter plot of the SXR
peak flux versus the HXR fluence for the complete sample, clearly
revealing an increase of $F_{\rm P,SXR}$ with increasing ${\cal
F}_{\rm HXR}$. It can also be inferred from the figure that the
slope is not constant over the whole range but that it is larger
for large HXR fluences than for small ones. We stress that
the slope at small fluences might be affected by missing events with
small SXR peak fluxes, and thus appear flatter than it is in fact.
The SXR threshold is basically caused by the GOES flare listings,
which generally do not embrace flares weaker than
B~class\footnote{Defined by a SXR peak flux without background
subtraction of $10^{-7}$~W~m$^{-2}$.}. The HXR fluence cut-off
arises due to the sensitivity limits of the HXR detectors as well
as due to constraints of the exact start/end time determination
for very short or weak flares.

\begin{figure}
\centering
\hspace*{-0.2cm}
\resizebox{0.95\hsize}{!}{\includegraphics{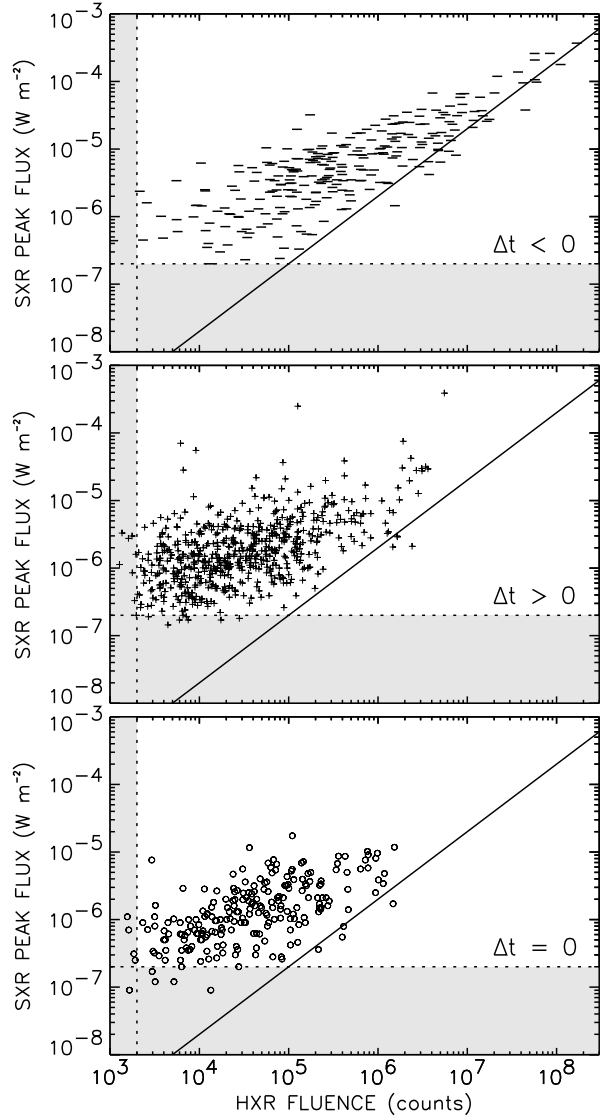}}
    \caption{Scatter plot of the SXR peak flux versus the HXR fluence
    separately for events with negative (top panel), positive
    (middle panel) and zero (bottom panel) time difference.
    The same line of constant~$k$ ($= 2\cdot 10^{-12}$) is shown in each case.
    \label{Fig_FluencePeak_sep} }
\end{figure}

In Fig.~\ref{Fig_FluencePeak} (note that the same holds for Figs.~3--9)
we have indicated the estimated thresholds of event detection by grey shading,
in order to visualize which ranges might be biased
by cut-off effects. From this representation it is evident that for the
range ${\cal F}_{\rm HXR} \gtrsim 2\cdot 10^5$~counts, the
scatter plot is not biased by the sensitivity thresholds.

The sign of the time difference between SXR peak and HXR end time of an event
is indicated by different plot symbols. ``$+$" symbols represent events with
$\Delta t > 0$, ``$-$" symbols events with $\Delta t < 0$. Events
that do not show a distinguishable time difference, i.e. the SXR peak and
HXR end take place within 1~min, are indicated by ``$\circ$" symbols.
In each of the figures a line of constant~$k$ is overplotted,
choosing $k = 2 \cdot 10^{-12}$~W~m$^{-2}$~counts$^{-1}$, which
can be considered as an estimate of $k$ for the largest flares.

Figure~\ref{Fig_FluencePeak_sep} shows the SXR peak flux -- HXR
fluence relationship separately for the events with $\Delta t > 0$,
$\Delta t < 0$ and $\Delta t = 0$. The figure clearly reveals
an interdependence between the importance of an event and the sign of
the time difference. Basically all large flares belong to the
group of events with $\Delta t < 0$, i.e. the SXR peak occurs before the HXR end.
On the other hand, this group covers distinctly fewer weak flares than
the group of events with $\Delta t > 0$.

\begin{figure}
\centering
\hspace*{-0.4cm}
\resizebox{0.91\hsize}{!}{\includegraphics{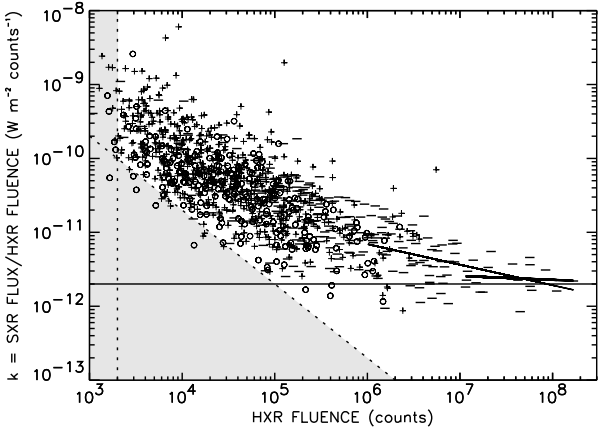}}
    \caption{Scatter plot of the factor~$k$, i.e. SXR peak flux divided
    by HXR fluence, versus HXR fluence for the complete sample.
    The same line of constant~$k$ ($= 2\cdot 10^{-12}$) is shown as in Fig.~2.
    Furthermore, we have plotted linear least-squares fits to the
    data, derived in the range ${\cal F}_{\rm HXR} > 10^6$~counts and
    ${\cal F}_{\rm HXR} > 10^7$~counts, respectively.
    \label{Fig_KHFluence} }
\end{figure}

From Figs.~\ref{Fig_FluencePeak} and~\ref{Fig_FluencePeak_sep}
it can be seen that, for very large fluences, the slope of the
SXR peak flux versus HXR fluence approaches the value of~1.
This phenomenon shows up even clearer in Fig.~\ref{Fig_KHFluence}, in
which the factor~$k$ as function of the HXR fluence is plotted.
In general, $k$ is decreasing for increasing ${\cal F}_{\rm HXR}$.
Yet, for large fluences, $k$ becomes nearly constant, indicating
an approximately linear relationship between the SXR peak flux and
the HXR fluence for the most intense events. We have applied a
linear least-squares fit to the data shown in Fig.~\ref{Fig_KHFluence} in the
range ${\cal F}_{\rm HXR} > 10^6$~counts and ${\cal F}_{\rm HXR} >
10^7$~counts. The obtained slopes give $b=-0.28\pm 0.06$ and $b=-0.05\pm 0.09$,
respectively. The least-squares fits are indicated in the figure by straight lines.

\begin{figure}
\centering
\hspace*{-0.4cm}
\resizebox{0.91\hsize}{!}{\includegraphics{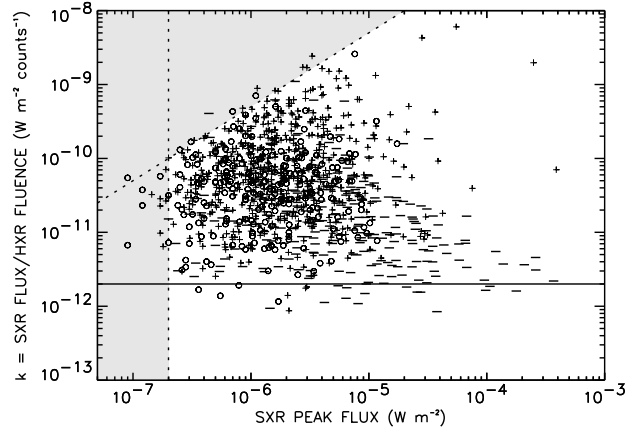}}
    \caption{Scatter plot of the factor~$k$ versus SXR peak flux for the
    complete sample. The same line of constant~$k$ ($= 2\cdot 10^{-12}$)
    is shown as in Fig.~2.
    \label{Fig_KSPeak} }
\end{figure}

In Fig.~\ref{Fig_KSPeak}, $k$ as function of the SXR peak flux is
plotted, revealing a very large scatter and a less distinct behavior
than for the HXR fluence. However, for events with negative time
differences (indicated by ``$-$" symbols), $k$ shows a tendency to decrease
with increasing SXR peak flux, approaching an almost constant $k$ for
very large peak fluxes (note that this phenomenon is not biased by
threshold effects). Events with positive time difference reveal a tendency
to increase with increasing SXR peak flux. However, it cannot be excluded
that this increasing behavior is biased by selection effects, missing
events with small SXR fluxes.

We obtain a high cross-correlation coefficient, $r=0.71$, for the
SXR peak flux and HXR fluence relationship. (All cross-correlation
coefficients are determined in log-log space). This coefficient is
higher than those for the SXR peak flux and HXR peak flux,
$r=0.57$. This indicates that the correlation is primarily due to the
HXR fluence -- SXR peak flux relationship, as predicted from the
Neupert effect, and not, e.g., due to the fact that flares
with high HXR peak fluxes also tend to have intense SXR counterparts.
However, since the fluence of an HXR event can be approximated by
the product of peak flux and event duration (e.g., Lee et al. 1995),
it is expected that the SXR peak flux -- HXR fluence correlation is
reflected also in a distinct correlation among the SXR and HXR peak
fluxes. The derived cross-correlation coefficients are very similar
to those reported by Wu et al. (1986) and Starr et al. (1988), who
analyzed selected samples of HXR/SXR bursts.

Finally, it is important to note that the HXR fluence~--
SXR peak flux correlation is higher for the
events with negative time differences, $r = 0.82$, than for the
events with positive time differences, $r = 0.54$ (see also Table~1).

\subsection{SXR peak flux -- HXR fluence analysis combined with
 the relative timing \label{Sect_combi} }

On the basis of the relative timing of the SXR peak and the HXR
end, we extracted two subsets of events. The events of set~1 are roughly
consistent with the timing expectations from the Neupert effect,
and the events of set~2 are inconsistent with it.
The two sets are defined by the following conditions:
\begin{eqnarray*}
{\rm Set~1:~} (|\Delta t| < 1 {\rm ~min}) & {\rm OR}  & (|\Delta
t_{\rm norm}| < 0.5{\rm ~unit}) \, ,\\
{\rm Set~2:~} (|\Delta t| > 2 {\rm ~min}) & {\rm AND}  & (|\Delta
t_{\rm norm}| > 1.0 {\rm~unit})\, .
\end{eqnarray*}
Out of the 1114 corresponding HXR and SXR flares, 485 (44\%) events fulfilled
the timing criterion of set~1; 270~events (24\%) belong to set~2; 359 events
(32\%) are neither attributed to set~1 nor to set~2.

The applied conditions represent a combination of absolute and
normalized time differences in order to avoid as much as possible
any a priori interdependence with the flare duration and/or flare
intensity. For example, the first part of the condition defining set~1,
which is based on absolute differences, is most likely to be
fulfilled in flares of short duration. On the other hand, the second part,
based on normalized differences, checks for consistency with the
Neupert effect in the case of long-duration flares. The exact
values chosen (1~min, 2~min; 0.5~unit, 1.0~unit) are, of course,
somewhat arbitrary. However, we stress that the two sets are
defined as non-adjoining, i.e. there is a significant fraction of
events (about one third) that are not used.  We have also
repeated the respective parts of the analysis with modified
values. This changed the number of events covered by the
respective sets but the results were qualitatively the
same, as long as the applied changes were not too large and the
data sets were not reduced too much.

\begin{figure}
\centering
\hspace*{-0.2cm}
\resizebox{0.95\hsize}{!}{\includegraphics{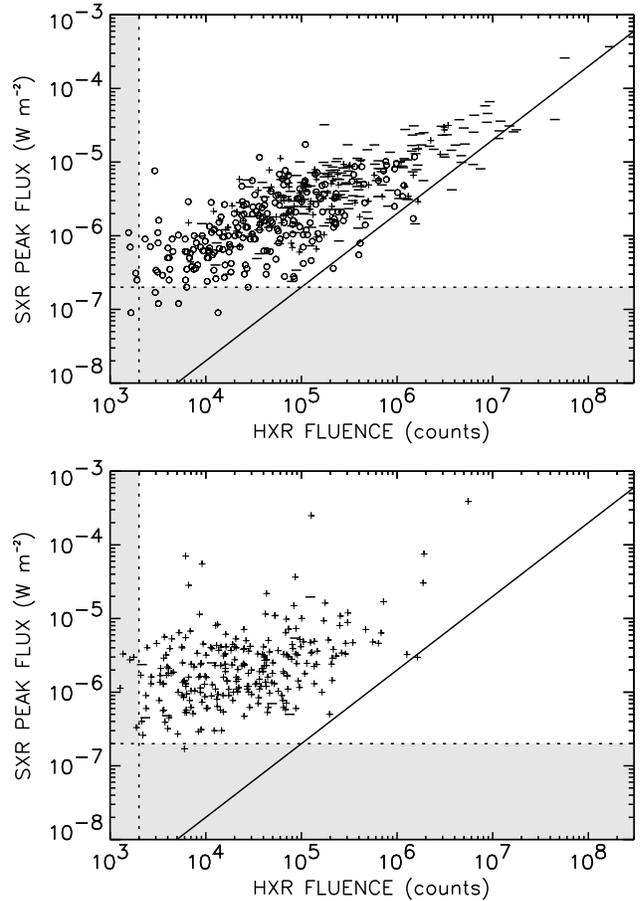}}
    \caption{Scatter plot of the SXR peak flux versus the HXR fluence
    separately plotted for set~1 (top panel) and set~2 (bottom panel).
    \label{Fig_FluencePeak2} }
\end{figure}

In the following, we analyze in detail the HXR fluence~-- SXR peak
flux relationship separately for both subsets in order to find out
whether there are distinct differences besides the temporal behavior.
Figure~\ref{Fig_FluencePeak2} shows the scatter plot of
the SXR peak flux versus the HXR fluence separately for set~1 and
set~2. Indeed, the two sets reveal very different characteristics.
Set~1 contains many more large events and shows a steeper increase
of $F_{\rm P,SXR}$ with increasing ${\cal F}_{\rm HXR}$ than set~2.
Moreover, set~1 contains many more events with negative than
positive time difference, although their absolute number is
much smaller (cf. Table~1). The subdivision of set~1 with regard to the time differences gives:
$\Delta t < 0$: 36\%, $\Delta t > 0$: 18\%, $\Delta t = 0$: 46\%.
Almost all events of set~2 (97\%) are characterized by $\Delta t > 0$,
i.e. increasing SXR emission while the HXR emission already stopped.
Obviously, considering the chosen criterion, it is not expected that
set~2 contains events with $\Delta t \le 0$
(see also the discussion in Sect.~\ref{Disc_Timing}).

Furthermore, for small fluences, the events belonging to set~2 have a
larger SXR peak flux at a given HXR fluence than do those of
set~1, indicating an ``excess" of SXR emission with respect to set~1.
For instance, the median of the SXR peak flux, determined in the range
${\cal F}_{\rm HXR} < 2\cdot 10^4$~counts, gives
$(6.5\pm 1.0)\cdot 10^{-7}$~W~m$^{-2}$ for the events of set~1, and
$(16.0\pm 3.2)\cdot 10^{-7}$~W~m$^{-2}$ for the events of set~2.
Note that this difference is not biased due to missing events
below the thresholds, as this should affect both sets in the same way.

\begin{table}
\caption{Cross-correlation coefficients derived for the SXR peak flux
and HXR fluence, and the SXR peak flux and HXR peak flux. The
correlations ($r$) are listed  for the total of events as well as
separately for events belonging to set~1/set~2 and events with
negative/positive time difference.}
\centering
\renewcommand{\arraystretch}{1.2}
\renewcommand{\tabcolsep}{1.5mm}
\begin{tabular}{lrrrrr} \hline
                                             & All  & Set 1  & Set 2 & $\Delta t$$<$0 & $\Delta t$$>$0\\ \hline
No. of events                                & 1114       & 485  & 270   & 269  & 622 \\
(\%) of total                                & 100        & 44   & 24    & 24 & 56  \\
$r\,(F_{\rm P,SXR}$ vs. ${\cal F}_{\rm HXR}$)  & 0.71     & 0.78  & 0.41  & 0.82 & 0.54\\
$r\,(F_{\rm P,SXR}$ vs. $F_{\rm P,HXR}$)       & 0.56     & 0.58  & 0.28  & 0.68 & 0.36 \\ \hline
\end{tabular}
\end{table}

The cross-correlation coefficients derived separately for the subsets
reveal that the correlation among the
SXR peak flux and HXR fluence is much more pronounced for the
events of set~1, $r=0.78$, than those of set~2, $r=0.41$.
A similar trend holds also for the SXR and HXR peak flux relationship.
The results of the cross-correlation analysis are summarized in Table~1.

\begin{figure}
\centering
\hspace*{-0.2cm}
\resizebox{0.93\hsize}{!}{\includegraphics{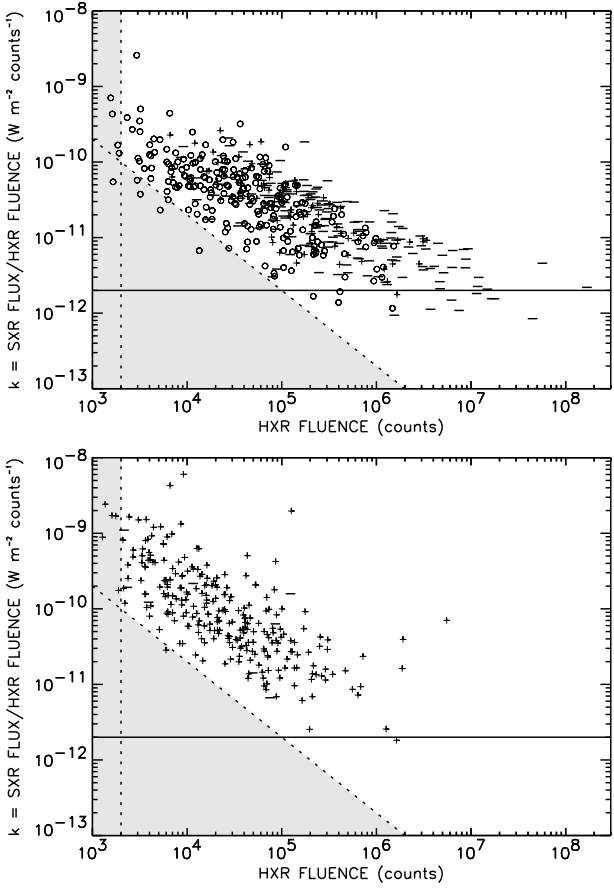}}
    \caption{Scatter plot of the factor~$k$ versus the HXR fluence
    for set~1 (top panel) and set~2 (bottom panel).
    \label{Fig_KHFluence2} }
\end{figure}

\begin{figure}
\centering
\hspace*{-0.2cm}
\resizebox{0.93\hsize}{!}{\includegraphics{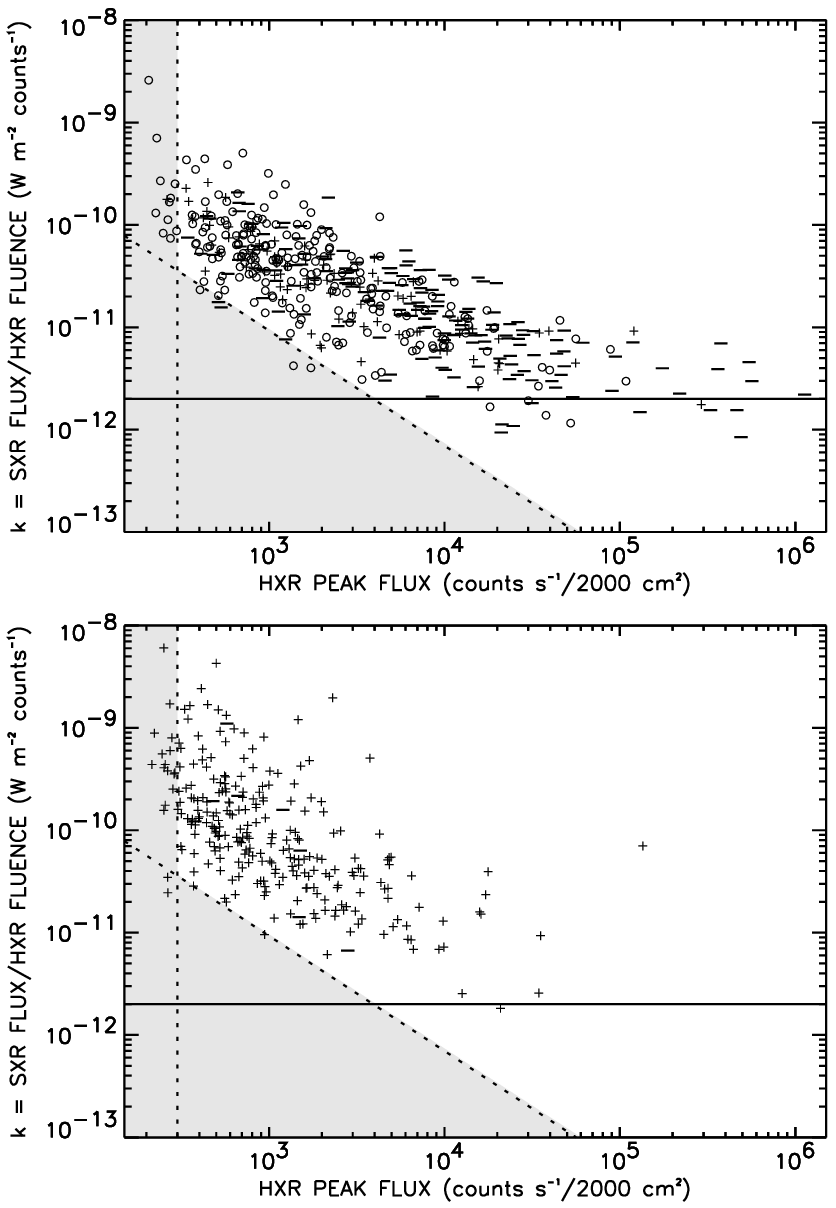}}
    \caption{Scatter plot of the factor~$k$ versus the HXR peak
    flux for set~1 (top panel) and set~2 (bottom panel).
    \label{Fig_KHPeak2} }
\end{figure}

\begin{figure}
\centering
\hspace*{-0.2cm}
\resizebox{0.93\hsize}{!}{\includegraphics{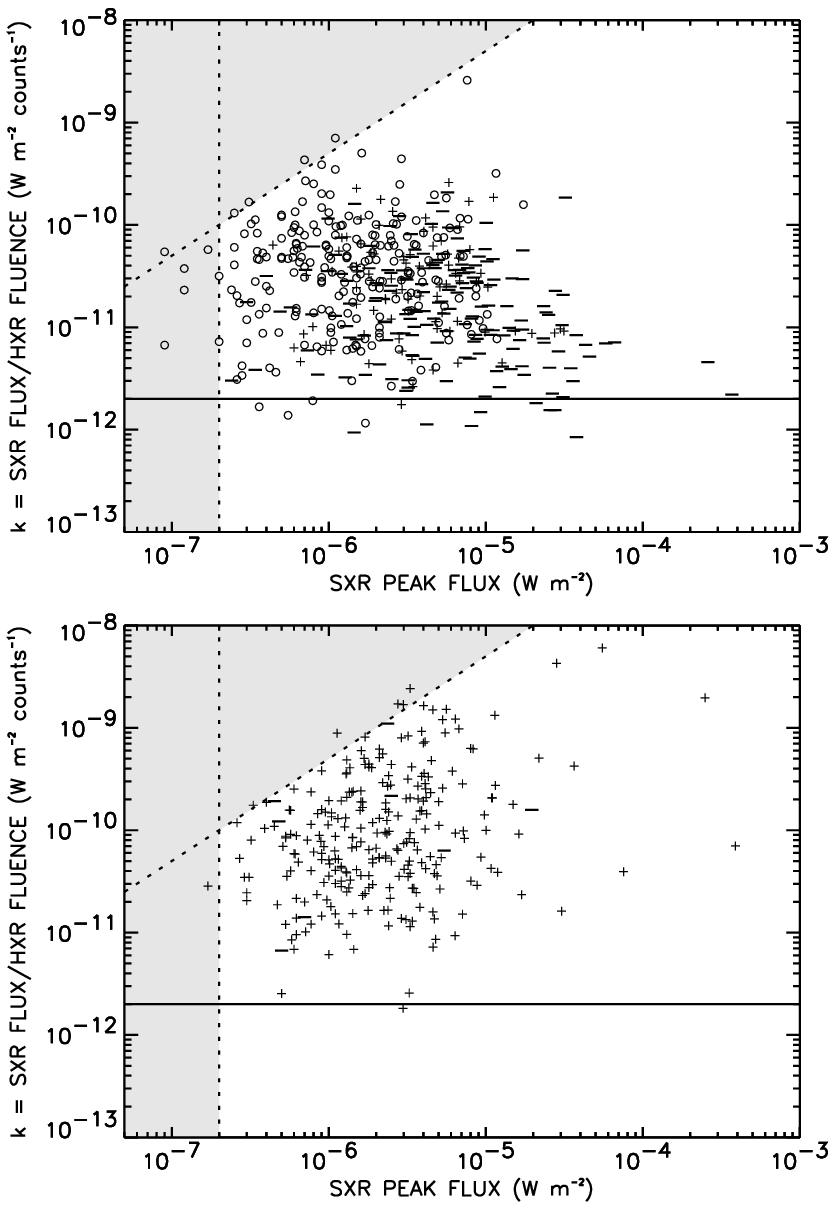}}
    \caption{Scatter plot of the factor~$k$ versus the SXR peak flux for set~1
    (top panel) and set~2 (bottom panel).
    \label{Fig_KSPeak2} }
\end{figure}

In Figs.~\ref{Fig_KHFluence2}, ~\ref{Fig_KHPeak2}
and~\ref{Fig_KSPeak2}, the factor $k$ is plotted as function of
the HXR fluence, the HXR peak flux and the SXR peak flux,
respectively. Fig.~\ref{Fig_KHFluence2} reveals that $k$ is a
distinct function of the HXR fluence. However, the specific
behavior is different for the two sets.  For set~2, the mean $k$ is a
rather monotonically decreasing function of the HXR fluence, with
a slope $b \sim -1.1$. For set~1, $k$ shows a different
behavior. For small and medium fluences, $k$ decreases with a
slope smaller than that of set~2, \mbox{$b \sim -0.8$}; for large
fluences, $k$ becomes almost constant, i.e. it is
approximately independent of the HXR fluence.

A quite similar overall behavior is found for the dependence of
$k$ on the HXR peak flux (Fig.~\ref{Fig_KHPeak2}), revealing also
a very distinct relationship between these parameters. In
Fig.~\ref{Fig_KSPeak2}, $k$ as function of the SXR peak flux is
plotted. Both sets reveal a large scatter. For set~1, a poor
anti-correlation exists between these two quantities, indicating
that $k$ only weakly depends on the SXR peak flux ($r\sim -0.2$).
For set~2, $k$ reveals a weak positive correlation with
the SXR peak flux ($r\sim 0.3$). However, this increase of $k$
with increasing SXR peak flux might be an artifact caused by threshold
effects.

We have applied least-squares fits in the form $\log (k) = a + b\cdot \log (A)$
for various SXR and HXR parameters, denoted here as $A$. Since we
are interested in the functional dependence between $k$ and the respective
HXR and SXR parameters, both variables should be treated symmetrically in the
fit procedure. This symmetric treatment is not satisfied by ordinary
least-squares regression of an dependent variable $y$ on an independent
variable~$x$. Isobe et al. (1990) have shown that regression using the
bisector of the two ordinary least-squares regression $y$ on $x$ and $x$ on
$y$ is the most suitable fitting method, if the goal is to determine the
underlying functional relationship between the variables. Thus, we
have applied the bisector fitting technique (for details see Isobe et al.\ 1990).

\begin{table*}
\centering
\caption{
Outcome of the regression analysis of $k$ as function of the HXR fluence,
the HXR peak flux and the SXR peak flux, determined separately for both sets.
The fits are of the form $\log(k) = a + b\cdot \log (A)$. Additionally, the
respective cross-correlation coefficients~$r$ are listed.}
\renewcommand{\arraystretch}{1.2}
\begin{tabular}{l|rrr|rrr|rrr} \hline
\multicolumn{1}{c}{$A$} & \multicolumn{3}{c}{${\cal F}_{\rm HXR}<3\cdot 10^5$} &
 \multicolumn{3}{c}{$F_{\rm P,HXR}<1\cdot 10^4$} &
 \multicolumn{3}{c}{$F_{\rm P,SXR}$} \\
\multicolumn{1}{l}{~} & \multicolumn{3}{c}{(counts)} &
\multicolumn{3}{c}{(counts s$^{-1}$\,/\,2000 cm$^2$)} &
 \multicolumn{3}{c}{(W m$^{-2}$)} \\ \hline
 &  \multicolumn{1}{c}{$a$} & \multicolumn{1}{c}{$b$} &  \multicolumn{1}{c|}{$r$} &
   \multicolumn{1}{c}{$a$} & \multicolumn{1}{c}{$b$} &  \multicolumn{1}{c|}{$r$} &
   \multicolumn{1}{c}{$a$} & \multicolumn{1}{c}{$b$}  & \multicolumn{1}{c}{$r$} \\
Set 1: & $-6.52$$\pm$$0.16$  & $-0.84$$\pm$$ 0.03$  & $-0.62$ &
        $-7.09$$\pm$$0.13$  & $-1.05$$\pm$$ 0.04$ & $-0.69$ &
        $-16.18$$\pm$$0.11$ & $-0.99$$\pm$$ 0.02$ & $-0.22$  \\
Set 2: & $-5.33$$\pm$$0.21$  & $-1.09$$\pm$$0.05$ & $-0.72$ &
        $-5.44$$\pm$$0.24$ & $-1.54$$\pm$$0.08$ & $-0.61$ &
        $-3.34$$\pm$$0.38$ & $+1.18$$\pm$$0.07$ & $+0.33$  \\ \hline
\end{tabular}
\end{table*}

The results of the regression analysis are summarized in Table~2. In
order to compare the respective fits of set~1 and set~2, we determined the
fits only in  a range of values of the respective parameters (indicated
in Table~2) in which both data sets cover a significant number of points.
The table reveals that the characteristics of both sets are
very different. In all cases, the difference of the fit parameters,
derived separately from set~1 and set~2 are larger than
the given uncertainties. (Note that due to the large scatter in the
graphs of $k$ versus the SXR peak flux (Fig.~\ref{Fig_KSPeak2}),
the respective fit parameters differ very much for different fitting
techniques, and have to be taken with caution.)

\section{Discussion}

\subsection{SXR -- HXR Timing \label{Disc_Timing}}

The timing analysis shows that 44\% of the events obey the
chosen criterion indicative of the Neupert effect, i.e. the difference
of the SXR peak time and the HXR end time is less than 1~min or less than
0.5~times the HXR duration, whereas 24\% reveal a strong deviation.
Obviously, this estimate depends on the chosen criteria. Nevertheless, the outcome
suggests that a large fraction of the events reveal a timing behavior that is consistent
with the Neupert effect; but there exists also a significant fraction of
events that are incompatible with the Neupert effect.

Comparing the distributions of the absolute and normalized time differences
(Fig.~\ref{Fig_timing}), it is noteworthy that a systematic difference shows
up for negative differences, denoting events for which the SXR maximum
occurs {\it before} the HXR end. Contrary to the distribution of the
absolute time differences, the distribution of the differences normalized
to the HXR duration reveals a sharp decline for negative differences,
indicating that a substantial fraction of the events with negative differences
are attributed to set~1. Indeed, 65\% of the events with $\Delta t < 0$ belong to
set~1, whereas only 14\% of the events with $\Delta t > 0$ do.

\begin{figure}
\centering
\hspace*{-0.1cm}
\resizebox{0.995\hsize}{!}{\includegraphics{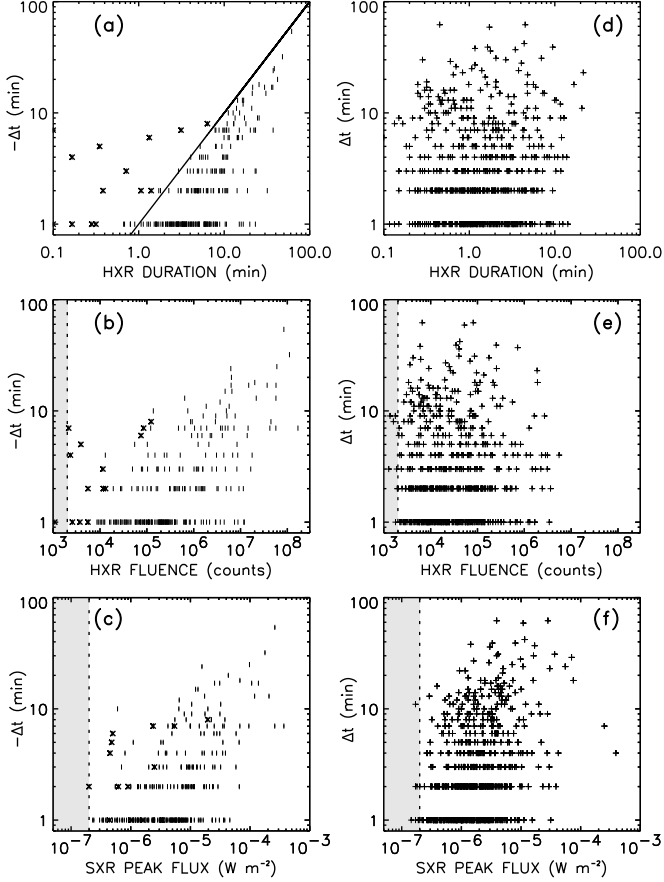}}
    \caption{We have plotted the absolute time difference between the
    peak of the SXR emission and the end of the HXR emission as a
    function of the event duration (top), HXR fluence (middle) and
    SXR peak flux (bottom), separately for the events with
    negative (panels {\bf a}--{\bf c}) and positive (panels {\bf d}--{\bf f})
    time differences. The line in panel~{\bf a} indicates
    $|\Delta t| = {\rm HXR}$ duration.
    \label{Fig_sign} }
\end{figure}

In Figure~\ref{Fig_sign}, we have plotted the absolute time difference
between the SXR peak and the HXR end as function of the event
duration, the HXR fluence and the SXR peak flux. Each of these
plots has been carried out separately for events with positive and
negative time differences, respectively. For the sample with negative
time differences, we have overplotted the line of $|\Delta t| = D$,
where $D$ is the HXR duration. For events that lie on the left hand side
of this line, $D$ is smaller than the absolute value of the time
difference, which can only mean that the SXR maximum
occurs before the start of the HXR emission. Taking into account the
given accuracy of 1~min for the SXR--HXR time differences,
14 events belong to this group (indicated by a
$\times$-symbol in panels~{\bf a--c} in Fig.~\ref{Fig_sign}). In principle,
there are two possible explanations for such events: either the respective SXR
and HXR events are not causally related but occur incidentally
within the applied 10-min start time window, or the given HXR duration
is estimated too short due to the sensitivity limits of the HXR detectors.
The second possibility is likely to apply for weak and short HXR events.

Comparing panels~{\bf a} and~{\bf d}, it is noticeable that the sample
with negative time differences covers very many events of
long duration, say, $D \gtrsim 10$~min
($\sim$70~events out of 269), whereas only a few events with long duration
belong to the sample with positive time differences ($\sim$15~events out
of 622). For the median HXR duration we obtain $\bar{D} = 5.3\pm 0.7$~min for the events with
$\Delta t < 0$ and $\bar{D} = 1.3\pm 0.1$~min for the events with \mbox{$\Delta t > 0$}.

Li et al. (1993) have calculated time profiles of spatially
integrated soft and hard X-ray emission from a thick-target
electron-heated model, finding that in general the time derivative
of the SXR time profile corresponds to the time profile of the HXR
emission, as stated by the Neupert effect. However, for long-duration
(``gradual") events they derived that this relationship
breaks down during the decay phase of the HXR event, in that the
maximum of the SXR emission occurs before the end of the HXR
event. This phenomenon can be explained by the fact that the SXR
emission starts to decrease if the evaporation-driven density
enhancements cannot overcome the cooling of the hot plasma, which
is likely to happen in gradual flares (Li et al. 1993).

Considering our observational findings together with the results from
simulations by Li et al. (1993), it is very likely that most of the events
with $\Delta t < 0$ are consistent with the electron-beam-driven evaporation model.
In particular, the very high correlation between the SXR peak flux and the HXR fluence
obtained for these events, \mbox{$r \sim 0.8$} (cf. Table~1), supports such interpretation.
From the histogram of the normalized time differences (cf.~Fig.~\ref{Fig_timing}, bottom panel),
we obtain a median $\overline{\Delta t}_{\rm norm} = -0.40\pm 0.04$ for the events
with $\Delta t < 0$. Thus, on average, the instantaneous cooling of the SXR emitting
plasma is dominating over the evaporation-driven energy supply for $\sim$0.4~times the HXR
duration during the decay phase of the HXR event, where the density is only slowly
increasing.

For the events with $\Delta t < 0$, there is a distinct correlation between the
absolute value of $\Delta t$ and the SXR peak flux
as well as the HXR fluence, $r \sim 0.6$ (cf. Fig.~\ref{Fig_sign},
panels~{\bf b} and {\bf c}). This correlation is basically caused by the
obvious correlation of the duration and the time difference
(cf. panel~{\bf a}), $r\sim 0.6$. For the events with $\Delta t > 0$,
we find no correlation between the absolute value of $\Delta t$
and the event duration as well as the HXR fluence, $r<0.1$ (cf. panels~{\bf d} and
{\bf e}). However, a weak but significant correlation exists between
$\Delta t$ and the SXR peak flux, $r=0.35$ (cf. panel~{\bf f}).
The fact that the time difference and the HXR fluence of an event are not correlated
but the time difference and the SXR peak flux are, might give indications
that for these events an additional energy transport mechanism other than the
HXR emitting electron beams is contributing to the SXR flux. 
On average, this contribution results in a comparatively higher SXR peak flux
for large time differences than small ones. Under this assumption, we can
estimate that, on average, the prolonged increase of the SXR emission takes
place for $\sim$2~times the HXR event duration ($\overline{\Delta t}_{\rm norm} = 2.00\pm 0.32$).

However, it has to be noted, that an extended heating beyond that
due to nonthermal electron beams is not the only possible explanation
for the events with $\Delta t > 0$. Using data from the Yohkoh Soft
X-ray Telescope (SXT) and Bragg Crystal Spectrometer (BCS),
McTiernan et al. (1999) have shown that consistency with the
Neupert effect depends on the SXR detector used. High-temperature
plasma ($T \gtrsim 16.5$ MK) is more likely than low-temperature
plasma to exhibit the Neupert effect. As shown by the authors, in
several flares, the thermal energy derived from the low-T
component was still increasing after the end of the HXR emission,
whereas the peak of the high-T component occurred almost
simultaneously with the HXR end. The curves of the total thermal
energy peaked somewhere in between those of the low and
high-T components. Thus, an increase of the emission of
low-T plasma after the HXR end does not necessarily indicate
an additional heating agent other than nonthermal electrons,
but alternatively it might be due to cooling of the high-T plasma
(see McTiernan et al. 1999). The GOES 0.1--0.8~nm detector used in
the present study also has a substantial response to low-T
plasma (Jim McTiernan, private communication). On the other hand, it
is more sensitive to high-T plasma than the SXT detectors used
in the study by McTiernan et al. (1999).

So, due to this arguments of McTiernan et al. (1999), we cannot simply attribute
all events with $\Delta t > 0$ as inconsistent with the Neupert effect and the
electron-beam-driven chromospheric evaporation model. Instead, we
consider as inconsistent only flares, which show strong deviations from
$\Delta t = 0$, i.e. the events belonging to set~2. From the present
study, we can infer that about half of the events fulfill the timing criterion
indicative for the Neupert effect (set~1), whereas about one fourth shows
strong violations of the Neupert timing with prolonged SXR increase after
the HXR end (set~2). This outcome is similar to the results from
McTiernan et al. (1999), applying a measure that correlates the derivative
of the SXR light curve with the HXR emission, that about half of the analyzed
events are compatible with the Neupert effect, whereas for the other
half an additional heating is indicated.

\subsection{Events of Set~2}

270 events ($\sim$25\% of the analyzed sample), belong to set~2,
i.e. they are characterized by strong deviations from the
timing expectations of the Neupert effect, with the SXR maximum occurring
distinctly {\it after} the HXR end. The fact that the SXR
emission is still increasing although the HXR emission, i.e. the
electron input, already stopped for more than 2~min and more than
1~unit of the HXR event duration, provides strong evidence that an
additional agent besides the HXR emitting electrons is contributing
to the energy input and prolonging the heating and/or evaporation. Possible
sources invoked as additional heating agents are, for instance, thermal
conduction (Zarro \& Lemen 1988; Yokoyama~\& Shibata 2001;
Czaykowska et al. 2001), accelerated protons (Simnett 1986;
Plunkett~\& Simnett 1994), plasma waves (Petrosian 1994; Lee et
al. 1995) and DC-electric fields (McDonald et al. 1999).

Comparing the HXR fluence -- SXR peak flux relationship for set~2 with
that for set~1 (Fig.~\ref{Fig_FluencePeak2}) clearly reveals that
both sets have very different characteristics. Set~2 has many
fewer large events, of, say, ${\cal F}_{\rm HXR} \gtrsim 2 \cdot 10^5$~counts,
than set~1. For very small fluences, on average, the events of set~2 have
larger SXR peak fluxes at given HXR fluences than those of set~1.
Moreover, for set~2 we obtain a distinctly smaller cross-correlation
coefficient between the HXR fluence and the SXR peak flux,
$r \sim 0.4$, than for set~1, $r \sim 0.8$ (cf. Table~1).
All these facts, the timing behavior, the higher SXR emission at a
given HXR fluence for weak flares, and a low correlation between
the HXR and SXR emission, provide evidence that the events of set~2,
comprising $\sim$25\% of the complete sample, are incompatible with the
scenario, in which the hot plasma giving rise to the SXR emission is
exclusively heated by thermalization of the electron beams responsible
for the HXR emission.

\subsection{Events of Set~1}

What about the events belonging to set~1? We cannot
straightforwardly conclude that these events are consistent with
the Neupert effect and the electron-beam-driven evaporation model.
The applied timing criterion presents a necessary condition for
the Neupert effect (which is definitely violated by the events of
set~2) but it is not a sufficient condition.  In principle, two
different possibilities can be distinguished. (i) The events of
set~1 obey the Neupert effect formulated for the energies
(Eq.~(\ref{NeupEn})), i.e. the electron-beam-driven evaporation model, but
there exists no simple linear proportionality between the X-ray
emissions, i.e. the factor~$k$ in Eq.~(\ref{EqNeup}) depends on
the flare intensity. In this case, $k$ contains information on the
relative productivity of soft X-ray emission per hard X-ray
emission as a function of flare importance. (ii) Not only in
the events belonging to set~2 but also in those of set~1, an
additional heating agent may be present. In this case, the heating
of the SXR plasma due to this additional agent must have
a similar timing to that of the nonthermal electrons; otherwise
the timing expectations of the Neupert effect would be
violated.

In the following section, we discuss the meaning of the
factor~$k$ from theoretical as well as observational
considerations, based on the presumption that the electron-beam-driven
chromospheric evaporation model is statistically valid for the
events of set~1 in the sense of possibility~(i), that they obey
the Neupert effect. In Sect.~5.5, the results will be discussed
with respect to possibility~(ii), i.e. under the presumption of an
additional heating agent.

\subsection{Relative productivity of HXR and SXR emission}

From theoretical considerations, it is evident that the relation
between the involved thermal and nonthermal energies and the
measured soft and hard X-ray emissions depend on a number of
parameters. This may introduce a systematic dependence of the SXR
peak flux -- HXR fluence relationship on the flare intensity.

The relation between the SXR peak flux, $F_{\rm P,SXR}$, and the
maximum thermal energy, $\epsilon_{\rm th,max}$, approximated by
the thermal energy at the time of the peak soft X-ray emission,
can be expressed as (cf. Lee et al. 1995):
\begin{equation}
\epsilon_{\rm th,max} \propto \frac{F_{\rm P,SXR} \, T_{\rm P}^{3/2}}{n_{\rm P}
I(T_{\rm P})} \, ,
\label{Eq_therm}
\end{equation}
where
\begin{equation}
I(T) = \int_{K_1/k_{\rm B}T}^{K_2/k_{\rm B}T} G(\xi,T)\,{\rm e}^{-\xi} d\xi \, .
\end{equation}
$T$ is the temperature, $n$ is the density, $k_{\rm B}$ is
Boltzmann's constant, and $K_1$ and $K_2$ are the energy limits of
the soft X-ray detector. The subscript~``P" indicates the value of
the respective quantities at the time of the peak SXR flux. Here
$G$ denotes the ratio of the actual emission at energy~$K$ to the
emission that would be appropriate for classical thermal
bremsstrahlung (for further discussion see Lee et al. 1995). The
relationship between the HXR fluence, ${\cal F}_{\rm HXR}$, and
the total energy deposited by nonthermal electrons,
$\epsilon_{{\rm e}^-}$, is given by (cf. Lee et al. 1995):
\begin{equation}
\epsilon_{{\rm e}^-} \propto {\cal F}_{\rm HXR} \left \langle
\gamma^2(\gamma -1) \left (\frac{E_0}{K_0}\right )^{1-\gamma}
\right \rangle \, , \label{Eq_el}
\end{equation}
with $E_0$ the low energy cut-off of the accelerated electron
spectrum, $K_0$ the photon energy, and $\gamma$ the photon
spectral index. The angle brackets denote the flux-averaging,
defined for a quantity $x(t)$ as:
\begin{equation}
\left \langle  x \right \rangle = \frac{\int_{t_0}^{t_0+D} x(t)F_{\rm HXR} (t)
\, dt}{
            \int_{t_0}^{t_0+D} F_{\rm HXR} (t) \, dt} \, .
\end{equation}

If the Neupert effect for the energies as expected from the
electron-beam-driven evaporation model obeys Eq.~(\ref{NeupEn}),
then from Eqs.~(\ref{Eq_therm}) and~(\ref{Eq_el}) it follows that the
factor~$k$, which relates the SXR peak flux and the HXR fluence
(Eq.~(\ref{EqNeup})), is given by:
\begin{equation}
k \propto \frac{n_{\rm P} I(T_{\rm P})}{T_{\rm P}^{3/2}} \left \langle
\gamma^2(\gamma -1)
\left (\frac{E_0}{K_0}\right )^{1-\gamma} \right \rangle \, .
\label{Eq_k}
\end{equation}
If the chromospheric evaporation model is valid for the considered
flare sample, $k$ gives the productivity of SXR emission relative
to the HXR emission. If any of the quantities on the right hand
side of Eq.~(\ref{Eq_k}) depends systematically on the flare
intensity, that must also be reflected in the behavior of $k$.

Feldman et al. (1996) found that the temperature of the SXR
emitting plasma shows a tendency to increase with flare intensity.
This would imply that $k$ should decrease with the importance of
the SXR burst. However, Garcia (2000) reported recently that, not
only the temperature but also the density of the SXR emitting
plasma tends to increase with increasing flare intensity. So, in
the first term on the right hand side of Eq.~(\ref{Eq_k}) the two
effects could, at least partly, compensate for each other. Such an
assumption is consistent with Fig.~\ref{Fig_KSPeak2} (top panel),
which shows that there is only a weak anti-correlation between $k$
and the SXR peak flux (see also Table~2, set~1).

All of this suggests that the second term on the right hand side
of Eq.~(\ref{Eq_k}), which is representative for the HXR emission,
might be responsible for a systematic dependence of $k$ on the
flare intensity.  This is also supported by the high anti-correlations
between $k$ and the HXR fluence as well as $k$ and the HXR peak
flux, in contrast to the poor anti-correlation between $k$ and the SXR
peak flux (cf. Figs.~\ref{Fig_KHFluence2}--\ref{Fig_KSPeak2}, top panels; Table~2, set~1).

Little is known about the cut-off energy~$E_0$, which in principle
may vary from flare to flare (e.g., Wu et al. 1986). For the
photon spectral index, $\gamma$, the situation is also rather
unclear. Datlowe et al. (1974), Dennis (1985) and Bromund et al. (1995)
did not find a correlation between the photon spectral index and the flare
intensity. However, from statistical considerations Lee et al.
(1995) inferred that a weak correlation between $\gamma$ and the
HXR fluence might exist. Tomczak (1999), who studied selected
flares observed by the Hard X-ray Telescope and the Soft X-ray
Telescope aboard Yohkoh, found that smaller values of $\gamma$
(harder HXR energy spectrum) tend to be associated with higher
hard X-ray intensities. Moreover, in this study it is reported that the
relative productivity of soft X-rays with regard to hard X-rays,
depends on the energy spectrum of the hard X-ray photons. A
steeper energy spectrum, i.e. larger $\gamma$, causes a higher
soft X-ray productivity with respect to the hard X-rays. In
principle, the results of Tomczak (1999) coincide with the findings of
the present study that the relative productivity of SXR emission
per HXR emission, i.e. $k$, is larger for weak HXR flares than for
intense ones (see Figs.~\ref{Fig_KHFluence2} and \ref{Fig_KHPeak2}, top panels).
However, it has to be stressed that the study of Tomczak (1999) is
based only on a sample of five events. Moreover, in this paper the hard
X-rays are primarily compared with SXR footpoint emission, which is only
a small fraction of the total SXR emission. Thus, this study is only of
limited relevance for the comparison of spatially integrated soft and
hard X-rays, as performed in the present paper.

On the basis of frequency distributions of HXR fluences and SXR
peak fluxes, Lee et al. (1995) derived a scaling of the SXR
parameters, i.e. $n_{\rm P}$, $T_{\rm P}$ and $I(T_{\rm P})$,
inferring the HXR fluence -- SXR peak flux relationship from the
differences of the power-law indices describing the SXR and HXR
frequency distributions. For the derivation of this scaling, two
assumptions were made: a) The electron-beam-driven evaporation model, i.e.
Eq.~(\ref{NeupEn}), holds for the bulk of flares; b) The
parameters that determine the relation between the total energy
deposited by electrons and the HXR emission, $\gamma$ and $E_0$,
do not systematically depend on the flare intensity, i.e.
$\epsilon_{{\rm e}^{-}} \propto{\cal F}_{\rm HXR}$, but only the
parameters that interrelate the thermal energy and the SXR
emission may vary with the flare intensity. As shown by the
authors, the derived scaling does not agree with observations.

The present study suggests that this discrepancy probably arises
due to inappropriate assumptions. We inferred that $\sim$25\% of
the events show strong deviations from the expectations of the
electron-beam-driven evaporation model. So the first assumption by
Lee et al. (1995) that the electron-beam-driven evaporation model
statistically holds for all flares is unlikely to be fulfilled.
Moreover, Figs.~7--9 (top panels) provide hints that the relation between SXR
emission and thermal energy does not systematically (or only weakly) depend on the
flare importance but the relation HXR emission -- nonthermal energy
does, just the opposite of the second assumption made by Lee et al. (1995).

Finally, let us note that flares are constituted of
multi-temperature plasma. McTiernan et al. (1999) have shown that
the high-temperature component is more likely to exhibit the Neupert effect than the
low-temperature component. In combination with the findings
of Feldman et al. (1996) and Garcia (2000) that the temperature increases
with the flare importance, this might imply that
more intense flares should show a more pronounced Neupert effect.
Indeed, Fig.~\ref{Fig_FluencePeak} shows that for intense flares the HXR
fluence~-- SXR peak flux relationship approaches a linear function.

\subsection{Interpretation in terms of an additional agent}

As mentioned in Sect.~5.3, the deviation from $k=const$ can be
also a result of an additional energy transport mechanism from the
energy release site. Since the energy release site is strongly
heated (see, e.g., Tsuneta 1996), a quite promising candidate for
such energy transport mechanism is thermal conduction (for
discussion see Vr\v{s}nak 1989; Somov 1992; McDonald et al. 1999;
and references therein), which is supported by various
observations. From high time resolution HXR and H$\alpha$
observations, K\"ampfer~\& Magun (1983) found evidence for the
occurrence of energy transport by electron-beams at one flare
kernel and for conductive energy transport at another kernel.
Moreover, various observations are indicative of prolonged
chromospheric evaporation driven by thermal conduction fronts
during the decaying phase of flares (e.g., Zarro \& Lemen 1988;
Czaykowska et al. 2001). Recently, extensive simulations of
chromospheric evaporation driven by thermal conduction have been
performed by Yokoyama \& Shibata (1998, 2001).

Under such an assumption of an additional heating agent, the
events of set 2 indicate that in a significant fraction of flares
this additional heat input can be even more important than that
associated with electron beams. The other extreme is where
electron-beam-driven evaporation is the dominant energy transport
mechanism, i.e. the second-agent contribution is negligible. However,
these flares are probably not the only constituents of set~1, but it
may embrace also flares, in which the energy deposition by high energy
electrons and those by the second agent have roughly the same time
evolution.

\begin{figure}
\centering
\hspace*{-0.25cm}
\resizebox{0.98\hsize}{!}{\includegraphics{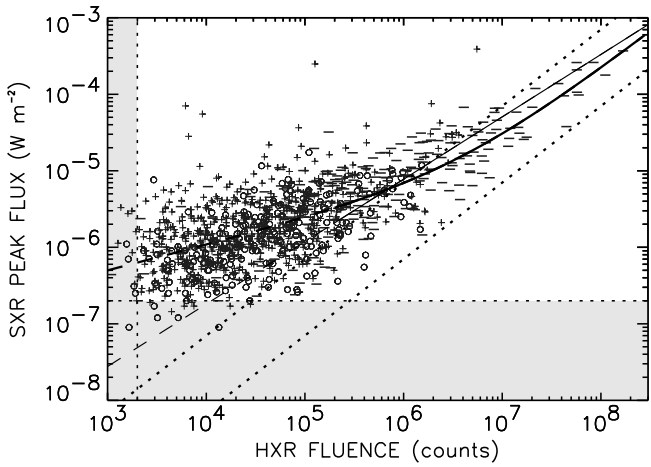}}
    \caption{Scatter plot of the SXR peak flux versus the HXR fluence
    for the whole sample. Several curves are overplotted:
    the functional fit according to Eq.~(\ref{k*}), represented by the curved
    line, and a linear fit (thin solid line), both derived for
    ${\cal F}_{\rm HXR} > 2\cdot 10^5$~counts, the extension of the
    fits to smaller fluences is indicated by dashed lines; two lines
    of constant~$k$, with $k=7 \cdot 10^{-12}$ (upper dotted line)
    and $k=0.7 \cdot 10^{-12}$ (lower dotted line).
    \label{Fig_fit} }
\end{figure}

To some degree, it is also expected that a Neupert-like relationship
between the hard and soft X-ray emissions shows up in thermal conduction
models (Dennis \& Schwartz 1989). For example,
it can be expected that a more powerful energy release results in
a higher temperature in the primary energy release site and thus a
higher thermal conduction flux. On the other hand, a powerful
energy release also implies higher electric fields in the primary
energy release site and thus a more efficient electron
acceleration. Under such circumstances, it can be assumed that the
relationship between $F_{\rm P,SXR}$ and ${\cal F}_{\rm HXR}$
has the form:
\begin{equation}
F_{\rm P,SXR} = \tilde{k} \cdot \,{\cal F}_{\rm HXR} + \tilde{k}^*
 \cdot ({\cal F}_{\rm HXR})^{\,\delta} \, .
 \label{k*}
\end{equation}
The first term on the right hand side represents the electron-beam-driven
contribution. The second term describes the contribution by
thermal conduction, which may (presumably weakly) depend on the
HXR flare importance. Here it is assumed that the coefficients $\tilde{k}$
and $\tilde{k}^*$ may differ from flare to flare depending on physical
conditions, but do not systematically vary with the flare
intensity. Under such an additional agent assumption,
set~1 and set~2 represent just the extremes of a continuum regarding
the two different types of energy input driving the evaporation, i.e. electron
beams and thermal conduction, and the whole sample of events should be
considered together.

We applied a least-squares fit in log-log space with the
functional form given in Eq.~(\ref{k*}), whereas only events
with ${\cal F}_{\rm HXR} > 2\cdot 10^5$~counts were taken into
account, as for this range any influence due to observational
thresholds can be excluded (cf. Fig.~\ref{Fig_FluencePeak}).
From the fit we obtained: $\tilde{k}=1.68\cdot 10^{-12 }$,
$\tilde{k}^*=1.27\cdot 10^{-8 }$ and $\delta = 0.45$.
We note that, qualitatively speaking, the fit yields a
similar result if it is applied to the events of set~1 only
($\tilde{k}=1.49\cdot 10^{-12 }$, $\tilde{k}^*=1.59\cdot 10^{-8}$,
$\delta = 0.44$). For comparison, we have also applied a linear fit
in log-log space, obtaining a slope $b = 0.83\pm 0.03$ in the range
${\cal F}_{\rm HXR} > 2\cdot 10^5$~counts. It is worth mentioning
that the linear fit for events with ${\cal F}_{\rm HXR} > 3\cdot 10^6$~counts
gives a slope of almost~1 ($b = 0.96\pm 0.07$), consistent with a linear
interpretation of the Neupert effect for these most intense events.
Figure~\ref{Fig_fit} shows the scatter plot of the SXR peak
flux versus the HXR fluence for the complete sample
(note that the data are identical to Fig.~\ref{Fig_FluencePeak})
in combination with the functional fit of Eq.~(\ref{k*}) as well
as the linear fit, derived from the events with
${\cal F}_{\rm HXR} \gtrsim 2\cdot 10^5$~counts.
The range, from which the fit was derived, is drawn in solid lines;
the extension of the fit to smaller fluences is represented by dashed lines.

As it can be seen from Fig.~\ref{Fig_fit}, Eq.~(\ref{k*}) approaches
a slope of~1 at large fluences. For small fluences, the fit probably
yields too high SXR peak fluxes, assuming that a considerable part of flares
lies below the indicated SXR threshold. According to the interpretation of
Eq.~(\ref{k*}) as a sum of two different energy contributions to
the evaporation, i.e. electron beams and thermal conduction, the derived
fit values imply that the electron-beam component is only dominating
for flares with ${\cal F}_{\rm HXR} \gtrsim 1.0\cdot 10^7$~counts
or, accordingly, $F_{\rm SXR,P} \gtrsim 1.4\cdot 10^{-5}$~W~m$^{-2}$.
These are rather improbably high values, as they imply that the electron-beam-driven
component is only dominating for the most intense events. However, we stress
that it is somewhat risky to apply such a functional fit to a limited
data set, and the derived fit values have to be taken with caution.
On the other hand, it has to be noted that the increasing slope indicated
by this functional fit is also reflected by the linear fit over
different fluence ranges. As shown above, for very large fluences
(${\cal F}_{\rm HXR} \gtrsim 3\cdot 10^6$~counts), the slope is
almost~1, whereas it is significantly lower if less intense
flares are also taken into account.

In Fig.~\ref{Fig_fit} we have also drawn a line of constant~$k$, with
$k = 0.7 \cdot 10^{-12}$. As can be seen from the figure, this value
represents a lower limit of~$k$. On the one hand, the line gives an impression,
in which range we might have missed events due to observational selection effects.
Furthermore, and more importantly, the existence of such a lower limit for~$k$
(which is quite sharply defined), contains very physical information:
electron beams always produce at a least a certain amount of soft X-rays
(see the ``white" triangle below the line $k = 0.7 \cdot 10^{-12}$).
There are no events with large HXR fluences but low soft X-ray emission.
On the other hand, the upper limit of $k$ is less well defined.
There are events with small HXR fluences but comparatively large SXR peak
fluxes, indicative for an additional agent besides the HXR emitting
electrons contributing to the energy of the SXR emitting plasma.
Moreover, the upper limit of~$k$ cannot be approximated
by a line of constant~$k$ (except, maybe, for flares with very large
fluences), but it reveals an increasing slope with increasing HXR
fluence. A similar shape can be detected in Fig.~5.4.2 from Wu et al.
(1986), who have drawn an analogous scatter plot for a selected sample of
101 events measured by the HXRBS and the BCS instrument aboard the
Solar Maximum Mission. It has to be noted that the shape of this upper
limit cannot be influenced by sensitivity effects, but it reflects a real
physical phenomenon, i.e. a different behavior of the SXR peak flux -- HXR
fluence relationship for small and large flares.

We have drawn also a line of
constant~$k$ in Fig.~\ref{Fig_fit} with $k=7 \cdot 10^{-12}$,
which represents an estimate of the upper limit of~$k$ for events
with ${\cal F}_{\rm HXR} \gtrsim 10^7$~counts. Thus, for the largest events,
the range between the lower and the upper limit of~$k$ covers about one
order of magnitude. Approaching smaller fluences, there is an increasing
number of flares lying above this range, i.e. at higher SXR
fluxes, violating the linear Neupert relation.

\section{Conclusions}

In the following we briefly summarize the basic results of the
analysis:
\begin{enumerate}
\item The distribution of the differences of the SXR peak times and
 HXR end times is strongly peaked at $\Delta t = 0$.
\item Yet, a significant fraction of events ($\sim$25\%)
 shows strong deviations from the chosen timing criterion applied as a
 necessary condition for consistency with the Neupert effect (set~2).
 These events are characterized by increasing SXR emission
 beyond the end of the HXR emission.
\item Flares that satisfy the timing criterion (set~1), embracing about
 one half of the events, reveal a much higher correlation between the
 HXR fluence and the SXR peak flux than those of set~2. The strong
 correlations found for set~1 suggests that electron-beam-driven
 chromospheric evaporation plays an important role for these events.
\item Set~2 contains many fewer large events than set~1. For weak flares, on
 average, the events of set~2 have higher SXR peak fluxes
 at a given HXR fluence than those of set~1, suggesting that an additional
 energy  transport mechanism other than the HXR emitting electrons
 contributes to the SXR emission.
\item Events with negative $\Delta t$, i.e. the SXR peak occurs
 before the end of the HXR emission, preferentially belong to
 set~1 and are of long duration. These events are compatible with
 the electron-beam-driven evaporation model. In the decay phase
 of long-duration flares, the instantaneous cooling of the hot plasma is likely
 to dominate over the evaporation-driven energy supply (Li et al.
 1993). From the present data set we infer that, on average,
 this phase covers $\sim$0.4~times of the HXR event duration.
\item For the events of set~1, the SXR peak flux -- HXR fluence
 relationship is not linear. However, for large HXR fluences the
 SXR peak flux -- HXR fluence relationship tends towards a linear
 function. Correspondingly, for the events of set~1, the factor $k$ is
 a decreasing function of the HXR fluence. Yet, for large HXR
 fluences, $k$ becomes approximately constant.
\item Although high correlations are found among the
 SXR peak fluxes and HXR fluences, the scatter in the SXR peak
 flux versus HXR fluence plot is larger than an order of magnitude
 (up to two orders of magnitude).
\end{enumerate}
Finally, we stress that although the results presented show that
in a statistical sense about half of the events show
characteristics compatible with the Neupert effect, the scatter of
the SXR peak flux versus HXR fluence indicates that a wide range
of physical conditions are met in solar flares. The main outcomes of the
analysis can be interpreted in the sense that the process of
electron-beam-driven evaporation plays an important role in solar flares.
On the other hand, the prolonged SXR emission found in a significant
fraction of events also gives strong indications for the presence of an
additional energy transport mechanism, probably thermal
conduction, whereas the relative contribution of the different transport
mechanisms shows a dependence on the flare importance. The energy
provided by the additional agent may play a prominent role in
weak flares, whereas in intense events its contribution is much less
important than the electron-beam-driven component.

\acknowledgements
The authors thank Helen Coffey from NGDC for making available the SXR
data, and the BATSE team for the access to the HXR data.
We also thank the referee Jim McTiernan for thoughtful comments.
A.\,V., M.\,T. and A.\,H. gratefully acknowledge
the Austrian {\em Fonds zur F\"orderung der wissenschaftlichen
Forschung} (FWF grants P13653-PHY and P15344-PHY) for supporting this project.
B.\,V. acknowledges the University of Graz for financial
support and is grateful to the colleagues from the Institute for
Geophysics, Astrophysics and Meteorology for their hospitality.

\end{document}